\documentclass[sigconf]{acmart}
\usepackage{booktabs} 

\setcopyright{acmcopyright}
\usepackage{float,graphicx}

\begin{document}

\copyrightyear{2018}
\acmYear{2018}
\setcopyright{acmcopyright}
\acmConference[PEARC '18]{Practice and Experience in Advanced Research Computing}{July 22--26, 2018}{Pittsburgh, PA, USA}
\acmBooktitle{PEARC '18: Practice and Experience in Advanced Research Computing, July 22--26, 2018, Pittsburgh, PA, USA}
\acmPrice{15.00}
\acmDOI{10.1145/3219104.3229281}
\acmISBN{978-1-4503-6446-1/18/07}


\title{Computational Challenges and Opportunities of \\ Simulating Cosmic Ray Showers at Global Scale}

\author{Olesya Sarajlic}
\affiliation{%
  \institution{Georgia State University  \\
                    Department of Physics and Astronomy
}
  \streetaddress{P.O. Box 5060}
  \city{Atlanta}
  \state{Georgia}
  \postcode{30302-5060}
}
\email{osarajlic@gmail.com}

\author{Xiaochun He}
\affiliation{%
  \institution{Georgia State University \\
                   Department of Physics and Astronomy
}
  \streetaddress{P.O. Box 5060}
  \city{Atlanta}
  \state{Georgia}
  \postcode{30302-5060}
}
\email{xhe@gsu.edu}

\author{Semir Sarajlic}
\affiliation{%
  \institution{Georgia Institute of Technology \\
                   Partnership for an Advanced Computing Environment
}
  \streetaddress{258 Fourth Street NW}
  \city{Atlanta}
  \state{Georgia}
  \postcode{30332-0700}
}
\email{ semir.sarajlic@oit.gatech.edu}

\author{Ting-Cun Wei}
\affiliation{%
  \institution{Northwestern Polytechnical University \\
                   School of Computer Science and Engineering
}
  \streetaddress{No. 1 Dongxiang Rd., Chang'an District}
  \city{Xi'an 710129, Shaanxi, PR}
  \country{China}
}
\email{weitc@nwpu.edu.cn}

\renewcommand{\shortauthors}{O. Sarajlic et al.}

\begin{abstract}

Galactic cosmic rays are the high-energy particles that stream into our solar system 
from distant corners of our Galaxy and some low energy particles are from the Sun which are associated with solar flares.  
The Earth atmosphere serves as an ideal detector for the high energy cosmic rays which interact with the air molecule nuclei causing propagation of extensive air showers. 
In recent years, there are growing interests in the applications of the cosmic ray
 measurements which range from the space/earth weather monitoring, homeland 
security based on the cosmic ray muon tomography, radiation effects on health via 
air travel, etc. 
A simulation program (based on the GEANT4 software package developed at CERN) 
has been developed at Georgia State University for studying the cosmic ray showers 
in atmosphere.  The results of this simulation study will provide unprecedented 
knowledge of the geo-position-dependent cosmic ray shower profiles and significantly
enhance the applicability of the cosmic ray applications. In the paper, we present the
computational challenges and the opportunities for carrying out the cosmic ray 
shower simulations at the global scale using various computing resources including XSEDE.

\end{abstract}

%
%



\maketitle

\section{Introduction}
\label{intro}
Galactic cosmic rays are the high-energy particles that stream into our solar system from distant corners of our Galaxy and some low energy particles are from the Sun which are associated with solar flares. 
The primary cosmic ray particles are mainly energetic protons ($>$79$\%$) and about 14$\%$ alpha particles, which are originated from supernovae explosions or other astrophysical events \cite{pdg:cosmic,dorman:crv}. The primary cosmic ray particles interact with the molecules in the atmosphere and produce showers of secondary particles (mainly pions) at about 15 km altitude. These pions are decaying into muons which are the dominant cosmic ray particle radiation (about 80$\%$) at the surface of the Earth.  

Over the past decades, numerous studies have reported the correlations between the dynamical changes of the Earth weather patterns and cosmic ray flux variation measured at the surface \cite{kirkby:climate,lu:correlation,ollila:changes,shaviv:climate}.  In recent years, other interesting applications of the cosmic ray measurements have been discovered which include the cosmic ray muon tomography for homeland security, volcanic activity monitoring, nuclear 
reactor core monitoring, etc. \cite{Pyramid_muon,Muon_tomography}.

The Nuclear Physics Group at Georgia State University (GSU) \cite{np6:phys} is currently developing novel, low-cost and portable cosmic ray detectors to be distributed around the world. One of the main goals of this project is to measure the cosmic ray radiation at the surface of the earth simultaneously at global scale to study the dynamical changes of the upper troposphere and the lower stratosphere. The success of this global measurement could lead to an unprecedented and accurate weather forecasting system both in short- and long-term. There are two computing related challenges for this project.  One is the need to monitor and collect data from the cosmic ray detector nodes in a world-wide cosmic ray detector network. The other is the systematic simulation of cosmic ray shower development in the atmosphere with variable geomagnetic field and atmospheric air density. To address the second challenge, a GEANT4-based cosmic ray shower simulation (ECRS) \cite{sanjeewa:simulation} has been developed to model cosmic ray showers in the Earth's atmosphere. The results of this simulation study will provide unprecedented knowledge of the geo-position-dependent cosmic ray shower profiles and significantly
enhance the applicability of the cosmic ray applications.

The GEANT4 software package \cite{agostinneli:simulation,allison:geant4} is widely applied in the field of high energy, nuclear and accelerator physics, as well as in medical and space sciences. The main goal of the ECRS simulation is to perform an extensive study of solar, geomagnetic field, temperature, and barometric pressure effects on cosmic ray showers in the atmosphere. 

In this study, we discuss the computational challenges of tracking all produced particles in each event in the whole depth of the atmosphere and sampling many events to obtain the statistically meaningful results.  We compare the benchmarks of our analysis across the computing resources that were available to us which include desktop workstations, campus cluster at GSU, computing facility at Brookhaven National Laboratory (BNL), and the Pittsburgh Supercomputing Center (PSC) Bridges \cite{towns:xsede}. 


In the following sections, we present an overview of the ECRS simulation and the use of computing resources to achieve statistically meaningful results. The details of the ECRS simulation setup is given in Section \ref{G4}.  In Section \ref{xsede}, we show our results from running the simulation on various computing resources. Section \ref{results} outlines the preliminary results from this study and the scalability of the ECRS simulation using various computing resources.  A brief summary and outlook is given in Section \ref{summary}.
\section{ECRS Simulation Setup}
\label{G4}




The ECRS simulation includes a realistic implementation of atmospheric air composition and density according to the US Standard Atmospheric Model \cite{lide:crc} and a time-dependent geo-magnetic field due to the varying solar activity \cite{finlay:igrf,tsyganenko:model}.  The earth atmosphere in the ECRS model is divided into 100 atmospheric layers in order to properly parameterize the air density variation as a function of the altitude. The atmosphere air consists of 78.09$\%$ N$_{2}$, 20.95$\%$ O$_{2}$, 0.93$\%$ Ar, and 0.03$\%$ CO$_{2}$. The earth is represented by the 11-kilometer shell consisting of water material, which allows one to study the cosmic ray radiation level at the depth of the ocean.

\begin{figure}[htb]
\centering
\includegraphics[width=0.3\textwidth]{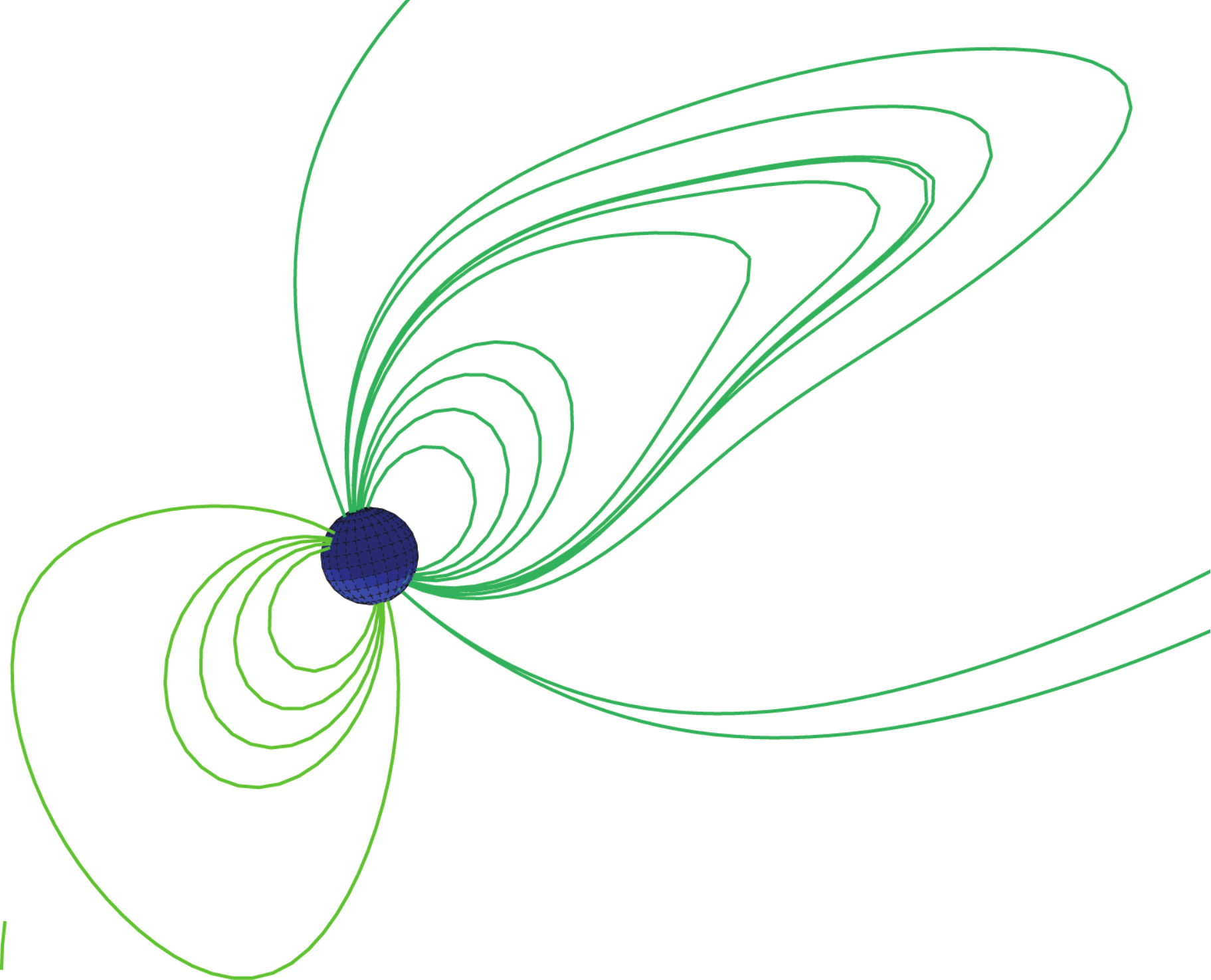} 
\caption{Visulization of the magnetic field lines around the Earth implemented in ECRS.}
\label{magneticFieldLines}
\end{figure}

The geo-magnetic field implemented in ECRS consists of the
internal and the external magnetic fields.  Internal geomagnetic field is given by the Internal Geomagnetic Reference Field model \cite{finlay:igrf}, and the external is using well established Tsyganenko models \cite{tsyganenko:model}.  Figure~\ref{magneticFieldLines} shows both, internal and external, field lines that are surrounding the Earth.  The internal field is fairly symmetric, while the sun facing side of the external is being compressed by the solar wind and the tail extends further in space. 

\begin{figure}
\centering
\includegraphics[width=0.4\textwidth]{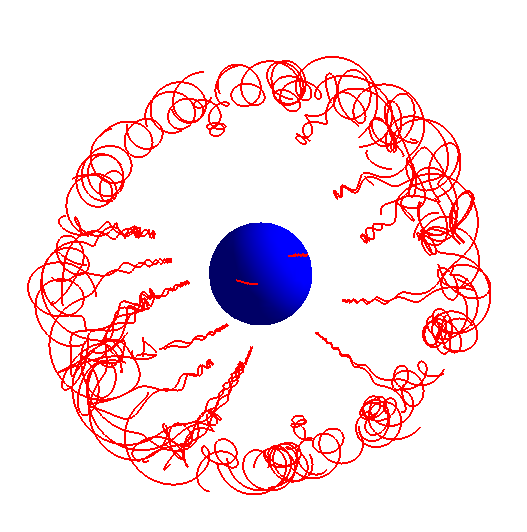}
\includegraphics[width=0.4\textwidth]{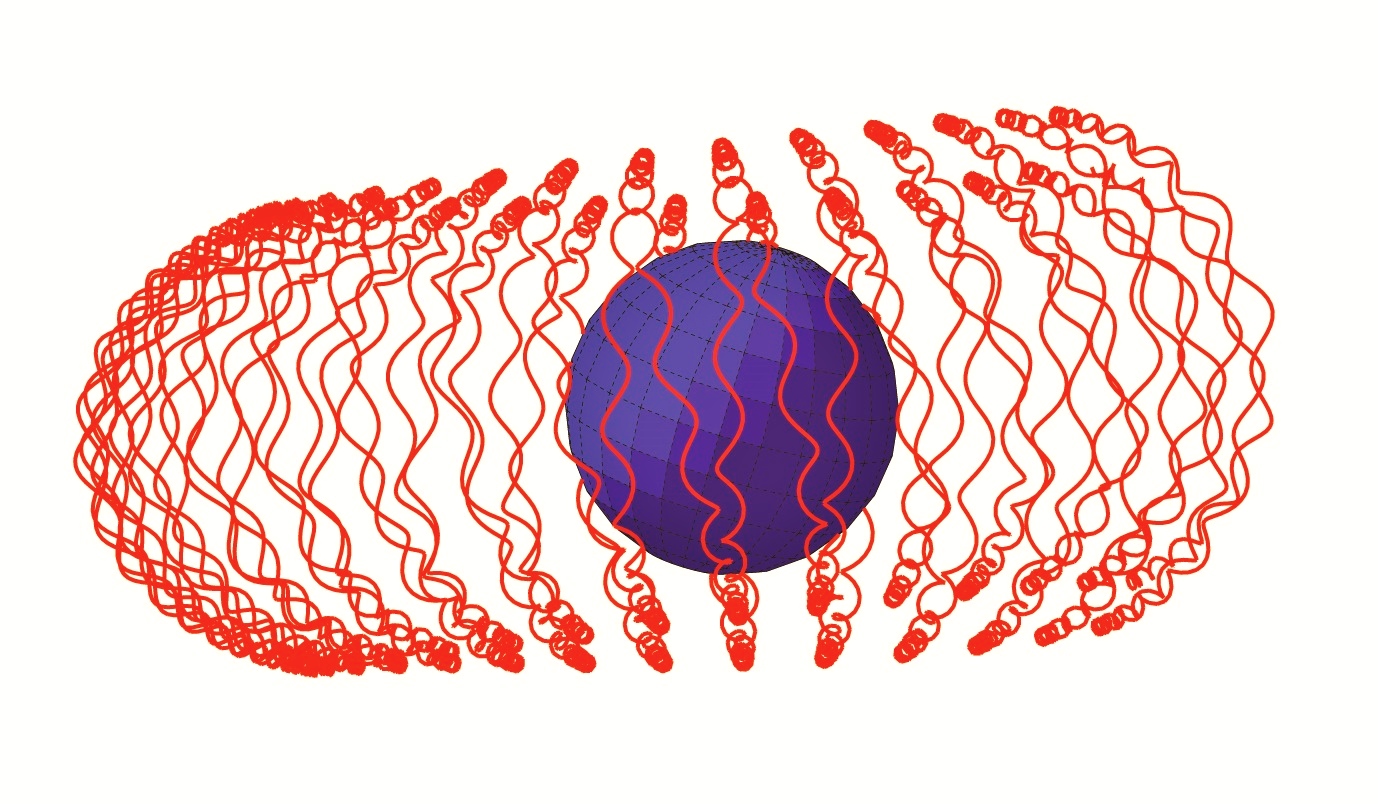} 
\caption{ (color online) The effect of geomagnetic field around the Earth. The motion of the proton on a magnetic shell is plotted in red, and the Earth is represented in blue.
Top panel: a top view visualization of the motion of 100 MeV protons on a geomagnetic shell during 60 seconds.
Bottom panel: a side view visualization of the motion of 10 MeV protons on a geomagnetic shell during 60 seconds.
}
\label{magcos_vis}
\end{figure}

To emphasize the complex structure of the magnetic field, Figure \ref{magcos_vis} shows its effect on the path of the low-energy incoming protons. 
This intricacy of the magnetic field impacts the computation time that is needed for tracking many of these particles in the simulation.

\begin{figure}[htb]
\centering
\includegraphics[width=0.40\textwidth]{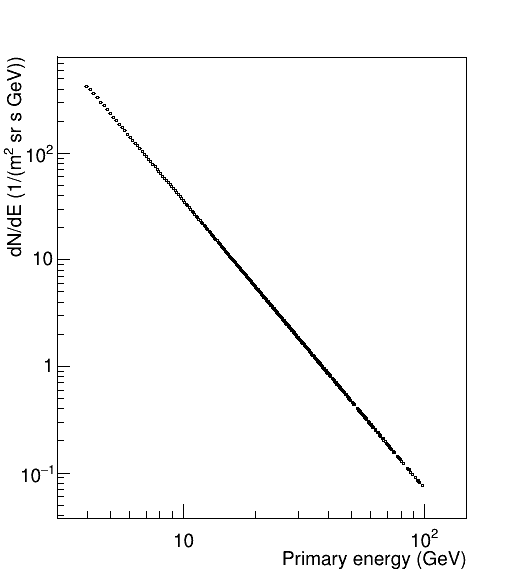} 
\caption{ (color online) Flux of protons of the primary cosmic rays in units of particles per energy as a function of energy.}
\label{Flux_eP_NP_noB}
\end{figure}

\begin{figure}[th]
\centering
\includegraphics[width=0.44\textwidth]{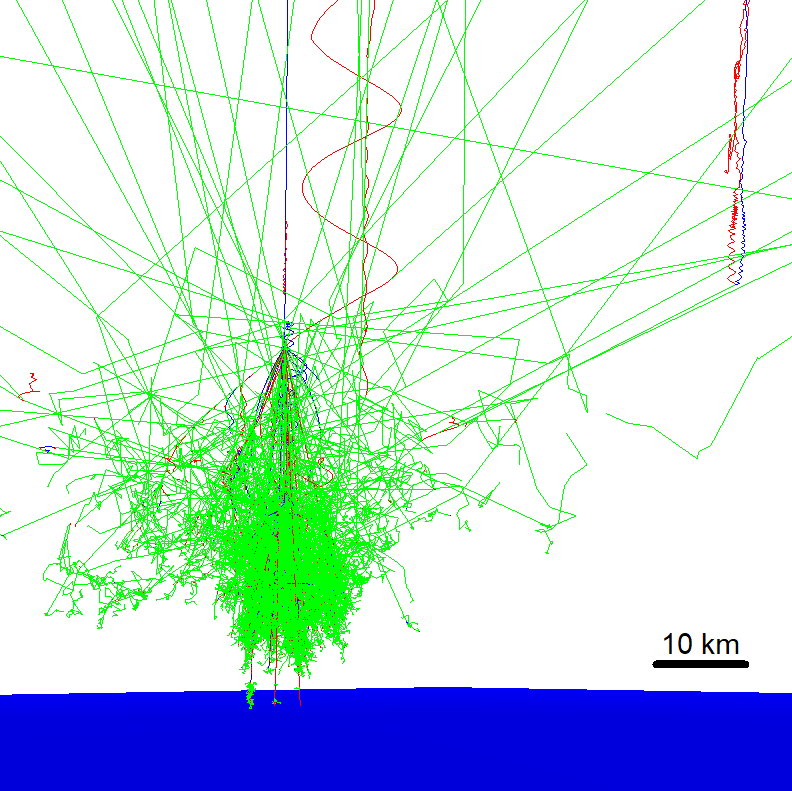}
\vspace{0.1in}
\caption{Cosmic ray shower event display from a 50 GeV primary proton launched toward the polar region.  The blue, red, and green color trajectories represent positive, negative, and neutral (gamma) particles, respectively. The curved trajectories are due to the magnetic field effect. 
}
\label{showerDevt}
\end{figure}

The primary cosmic ray particles that we are interested in studying in this project are protons with energies below 100 GeV which are dominant in the primary cosmic ray spectrum, as shown in Fig.~\ref{Flux_eP_NP_noB}.   
Figure~\ref{showerDevt} shows a cosmic ray shower event display from the ECRS simulation produced by a single 50 GeV proton. As shown in Fig.~\ref{showerDevt}, most of the secondary particles are produced around 15 km in altitude, which is a few km higher than the typical flight altitude of air travel. It is for this reason that the study of cosmic ray shower activities is also important for understanding the health hazard for flight crew.

A typical event, as shown in Fig.~\ref{showerDevt}, will take on average about 10 minutes to complete on a Linux desktop environment.  To carry out the cosmic ray shower simulation with statistically meaningful results is computationally demanding considering the following factors:
\begin{itemize}
\item Accumulating large number of cosmic ray shower events at a given geo-position
(i.e., geo-magnetic field variations) with variable atmospheric air density profile;
\item Tracking low energy cosmic ray shower particles at the earth-size scale (i.e., computing time consumption);
\item Outputting extensive shower particle information produced in the atmosphere for offline  data analysis.
\end{itemize}

ECRS simulation is intrinsically parallel at per event level.  This means that one could run events independently of one another on different compute resources.  ECRS is an exceptionally optimized code that utilizes 100$\%$ of the CPU throughout the duration of the event run, which results in a 100$\%$ resource utilization of the resource reserved via a workload scheduler.  In the following section, we demonstrate the scalability of ECRS simulation from personal computer to institutional small scale cluster and later national resources at XSEDE and BNL.

\section{Computational Challenges of Running ECRS Simulation}
\label{xsede}

\subsection{Desktop Computer}
\begin{figure}[htb]
\centering
\includegraphics[width=0.4\textwidth]{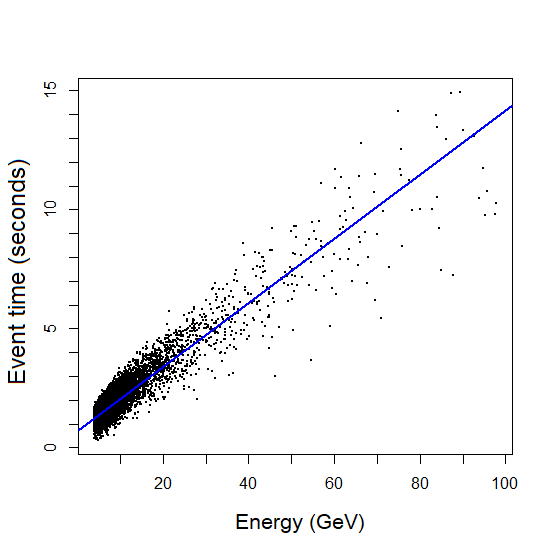} 
\includegraphics[width=0.4\textwidth]{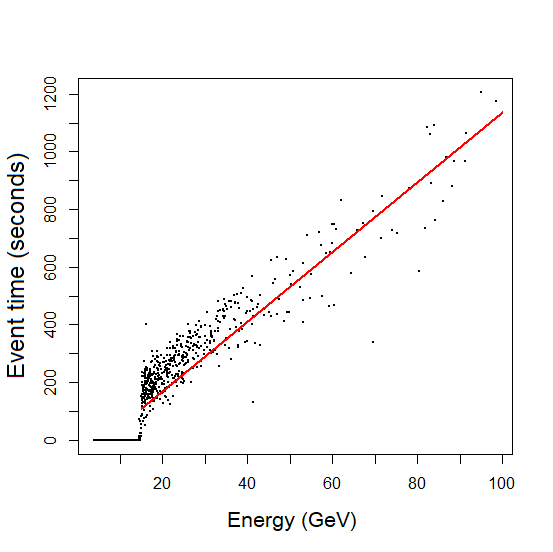} 
\caption{Scatter plot of event time vs. the incident primary particle energy. 
The cosmic ray particles are launched from 1.2 Earth's radii toward the center of the earth to
33.75$^\circ$ North and 264.39$^\circ$ East geo-position. Top panel: cosmic ray shower simulation without geo-magnetic field.
Bottom panel: cosmic ray shower simulation with geo-magnetic field.}
\label{time_ke}
\end{figure}

In order to provide a reference for assessing the computing resources in XSEDE, we ran ECRS on 
a high-end desktop machine (Mac Pro: 3.5 GHz 6-Core Intel Xeon E5 with 64 GB RAM) 
by launching cosmic rays toward 33.75$^\circ$ North and 264.39$^\circ$ East from 1.2 Earth's radius in altitude. 
The CPU execution time (i.e., event time) of this simulation per cosmic ray event as a function of the primary particle energy is shown in Fig.~\ref{time_ke} with and without geo-magnetic field. 
As it is expected, it takes much longer CPU time for tracking particles in the geo-magnetic field. For example, for a proton at 60 GeV energy, it only takes on average 9 seconds to complete the event without the geo-magnetic field in comparison to 700 seconds with the geo-magnetic field.

It is also interesting to notice here that it takes very little CPU time when the primary energies are 
less than 15 GeV in case when the geo-magnetic field is enabled.  In other words, the geo-magnetic field will deflect low-energy primary cosmic ray particles away from entering into the earth atmosphere. Given the fact that the geo-magnetic field is non-uniform and asymmetric, one needs to run ECRS simulation at each location accordingly in order to properly take into account the field effect. This ultimately brings the computing challenges to carry out these simulations with reasonably achievable statistical accuracy.

\subsection{GSU Cluster}
GSU's Orion \cite{sarajlic:orion} is a heterogeneous Linux cluster comprised of 360 cores, 4.25 TB of RAM, and 87 TB of NFS storage with 
LSF workload manager for scheduling batch and interactive jobs.  This is an ideal system for testing the ECRS simulation performance by submitting many batch jobs in parallel to multiple nodes in order to achieve higher statistics.  

\begin{figure}
\includegraphics[width=0.5\textwidth]{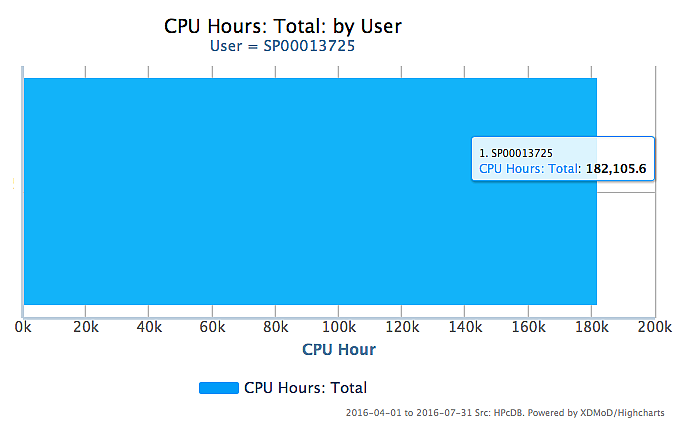}
\caption{ (color online)  
CPU hours used by Nuclear Physics Group/user (SP00013725) on Orion between 04/01/2016 and 07/31/2016, which is 182,105.6 CPU hours. 
}
\label{CPU_Orion}
\end{figure}

For this test, we used a total of 182,105 CPU hours between April, 2016 and July, 2016 as shown in Fig.~\ref{CPU_Orion} 
which accounted for 32.3$\%$ of overall cluster utilization during that period \footnote{Open XDMoD \cite{palmer:xdmod} 
for Georgia State University: {\url{http://xdmod.rs.gsu.edu}}}.  While this computing resource was not sufficient for achieving the required event statistics, we were able to successfully run our ECRS simulation in a shared 
cluster environment. We then turned to larger national resources to supplement our computing needs.

\subsection{RHIC Computing Facility}
\label{RHIC}
\begin{figure*}
\centering
\includegraphics[width=0.9\textwidth]{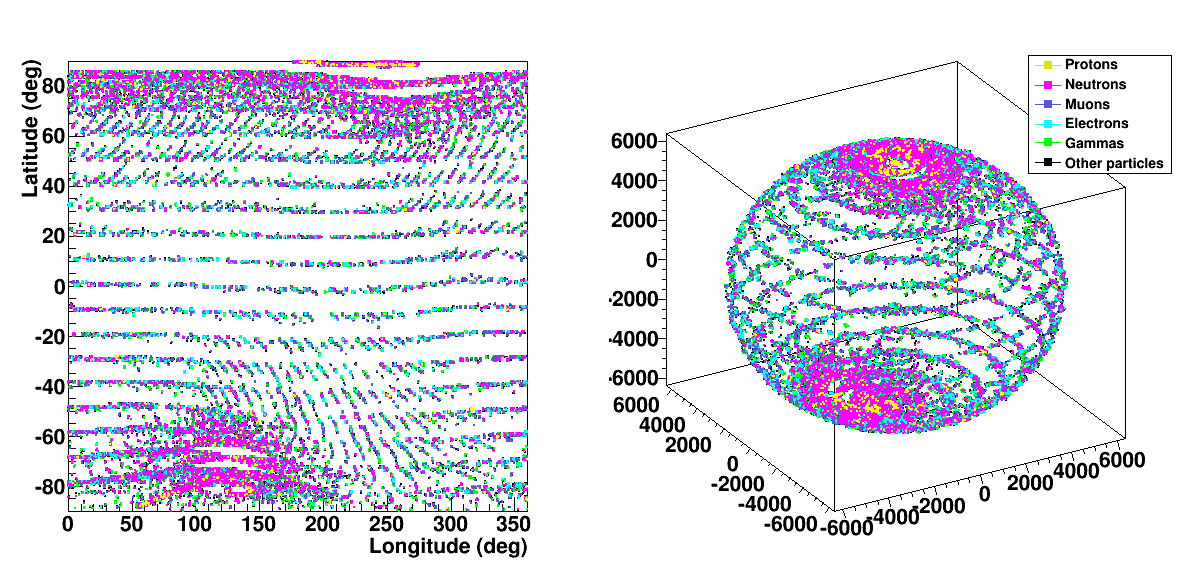} 
\caption{Left panel: Latitude versus longitude distributions of particles that reached the surface of the earth (with magnetic field being implemented). 
Right panel: 3D display of the global particle distributions at the surface with magnetic field implementation.}
\label{global_B}
\end{figure*}

In order to qualitatively explore the magnetic field effect on the cosmic ray shower development in the atmosphere, we ran the ECRS simulation on a computing farm (well-over 10,000 nodes) at the RHIC Computing Facility at BNL by launching primary cosmic rays from 1.2 Earth's radii toward the surface of the earth at 10 degree increment in latitude and longitude. This simulation exercise was divided into 6840 batch jobs and took more two weeks of time to complete 1000 events per batch job. 
Figure~\ref{global_B} shows the distributions of the ionizing particle radiation (including protons, neutrons, muons, electrons and gamma rays) that reached sea level. 
While it is very clear to see the geo-position dependent cosmic ray shower particle distributions at the surface of the earth as shown in Fig. \ref{global_B}, it is still statistically limited toward obtaining adequate distributions to quantify this variation in any of the interesting applications aforementioned. 

\subsection{XSEDE Computing Resources}

\begin{figure}
\centering
\includegraphics[width=0.5\textwidth]{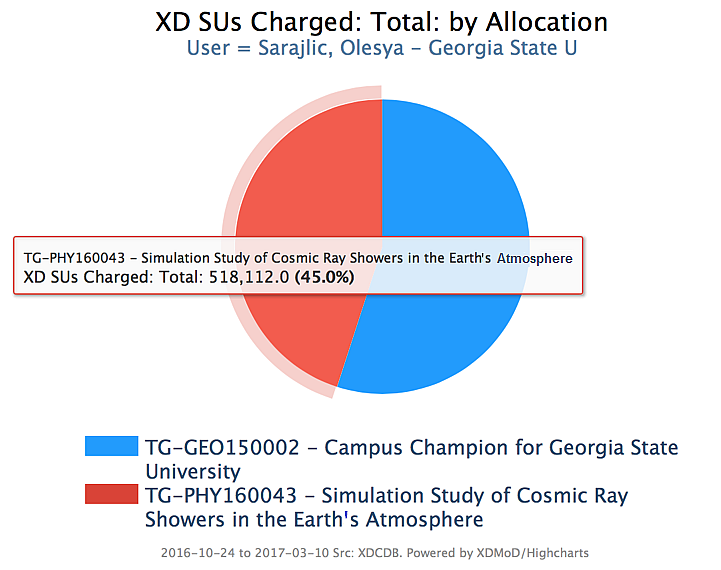}
\caption{ (color online) Service units charged from XSEDE grants by allocation: 55$\%$ from GEO150002 and 45$\%$ from PHY160043.}
\label{SUs_Charged_byAllocation}
\end{figure}

\begin{figure}
\includegraphics[width=0.5\textwidth]{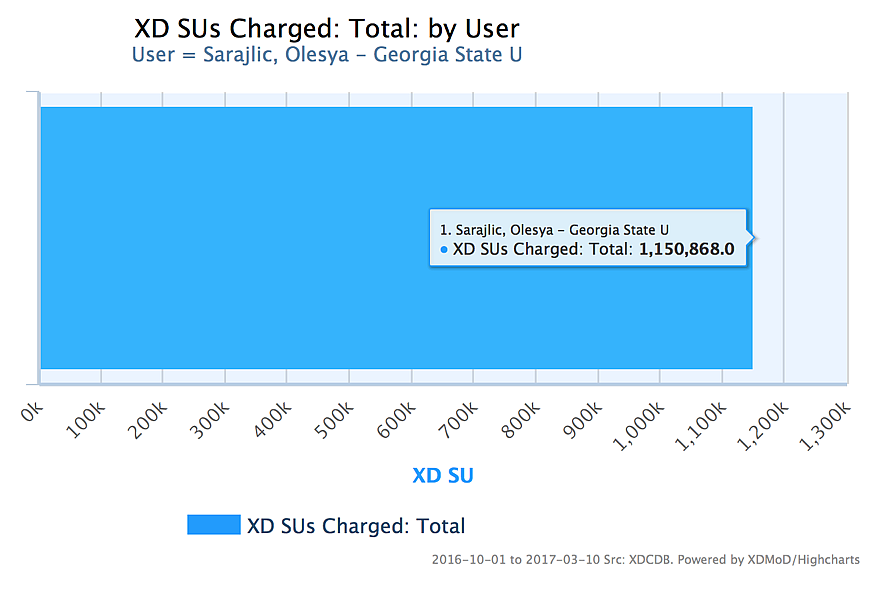}
\caption{ (color online)  
Service units charged by user on XSEDE between 10/01/2016 and 03/10/2017, which is roughly 200,000 CPU hours. 
}
\label{SU_CPU_byUser}
\end{figure}

Through XSEDE Campus Champion (GEO150002) and startup allocation (PHY160043) grants we got access to Bridges cluster at Pittsburgh Supercomputing Center (PSC). Figures \ref{SUs_Charged_byAllocation} and \ref{SU_CPU_byUser} show total SUs charged by allocation from both XSEDE grants. We ran our simulation on the Regular Shared Memory (RSM) computational nodes comprised of HPE Apollo 2000s with 2 Intel Xeon E5-2695 v3 CPUs (14 core per CPU),  128 GB RAM, and 8TB on-node storage. 

Similar to the ECRS simulation setup as described in Section \ref{RHIC}, we launched the primary cosmic ray particles over 4$\pi$ steradian in direction with 10 degree increments both in geographic latitude and longitude.  Given the extended CPU time needed per event with the geo-magnetic field on, each job is limited to 500 events in order to complete the batch jobs within the 48-hour requirement on the XSEDE Bridges.  
From 10/01/2016 to 03/10/2017, as shown in Figures \ref{SU_CPU_byUser}, we consumed a total of 1,150,868 SUs (186,675.9 CPU hours) for running these simulation jobs, which greatly exceeded the total SU allocation for the combined PHY160043 and GEO150002 projects. One of the major reasons of consuming a large number of SUs was related to the 48-hour requirement per single batch job. In some cases, if the sampled primary particle energy is very high, it takes tremendous amount of CPU time to track all low-energy secondary particles produced in the cosmic ray shower which in turn prevents the completion of the total number of events for the batch job.  If this happens, we had to re-submit the incomplete batch job in order to meet the required statistics at each geo-position, which we automated via a script that would scan the job outputs and resubmit the incomplete jobs from the position where the initial job completed.  

For achieving meaningful statistical accuracy, one needs to simulate  more than 10,000 events at each geo-position. This would require an XSEDE allocation greater than 2 million SUs for running ECRS simulation to accumulate statistically accurate results at the global scale. We will continue exploring the computing opportunities at XSEDE in our future work.

\section{Results and Discussion}
\label{results}

In this section, we highlight some of the important results of cosmic ray shower simulations that we could obtain by using the computing resources aforementioned. As it is shown in bottom panel of Fig.~\ref{time_ke}, more CPU time is required to run cosmic ray shower simulation for greater primary particle energies (approximately a linear relationship above the cutoff energy). 
As seen in Fig.~\ref{Flux_eP_NP_noB}, most of the cosmic ray shower events have lower energies that require less CPU time to complete the shower simulation. However, there is a small percentage of events which has tens of GeV energies and takes up most of the total CPU time of the simulation batch jobs. This is a trend that we experienced across the computing resources we studied so far.

\begin{figure}[htb]
\centering
\includegraphics[width=0.40\textwidth]{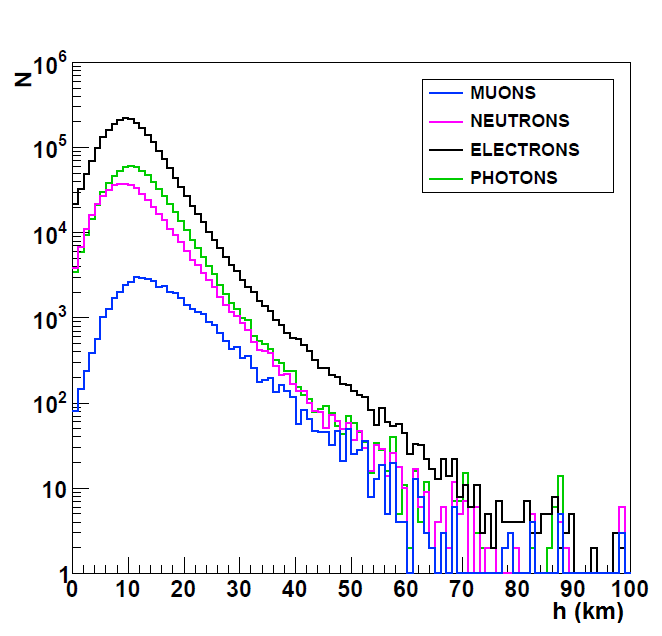} 
\caption{ (color online) Distributions of the secondary cosmic ray particles in the atmosphere as a function of the altitude: muons (blue curve), neutrons (magenta curve), electrons (black curve), and gamma ray photons (green curve). The color code is match the color scheme in Fig.~\ref{global_B}.}
\label{Altitude_NP_NoB}
\end{figure}

One of the innovative features of the ECRS simulation is that one can simulate cosmic ray shower activities simultaneously at global scale as shown in Fig.~\ref{global_B}. It is impossible to complete this task in a single or small cluster desktop environment. However, based on our initial test with national resources via BNL and XSEDE, it is very possible to carry out the ECRS simulations with large statistics if more resources can be allocated. As an example, one could track all particle species produced in the cosmic ray showers at each geo-position in the whole atmosphere, as shown in Fig.~\ref{Altitude_NP_NoB}. This study is important for using the cosmic ray muon and neutron particles to determine the effective atmospheric temperature in the higher altitude in atmosphere (>6 km).

\begin{figure}[htb]
\centering
\includegraphics[width=0.40\textwidth]{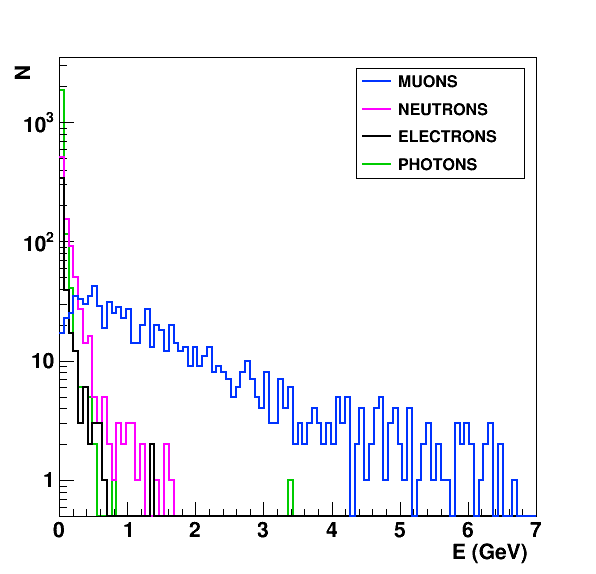} 
\caption{ (color online) Particle energy distributions of muons (blue curve), neutrons (magenta curve), electrons (black curve), and gamma ray photons (green curve). }
\label{KE_NP_NoB}
\end{figure}

The possibility of carrying out the ECRS simulation with large statistics also allows us to study the cosmic ray radiation budget (i.e. mainly muon and neutron particle flux) at the surface of the earth at global scale simultaneously. Figure \ref{KE_NP_NoB} shows the particle energy distributions of muons, neutrons, electrons and photons for a given geo-position as an example. This information is important since one could use it to compare with the results from the cosmic ray detectors installed at different locations at the surface of the earth\footnote{World Data Center for Cosmic Rays: \url{http://cidas.isee.nagoya-u.ac.jp/WDCCR/}}.

\begin{figure}[htb]
\centering
\includegraphics[width=0.41\textwidth]{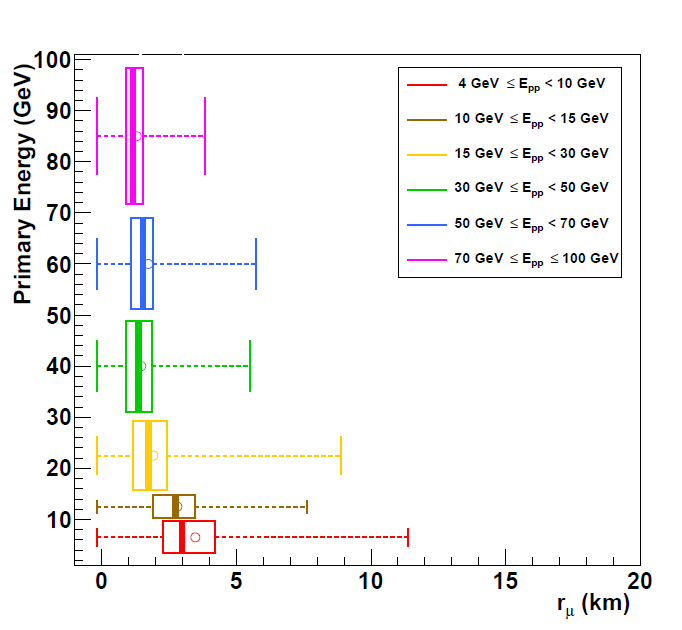} 
\includegraphics[width=0.4\textwidth]{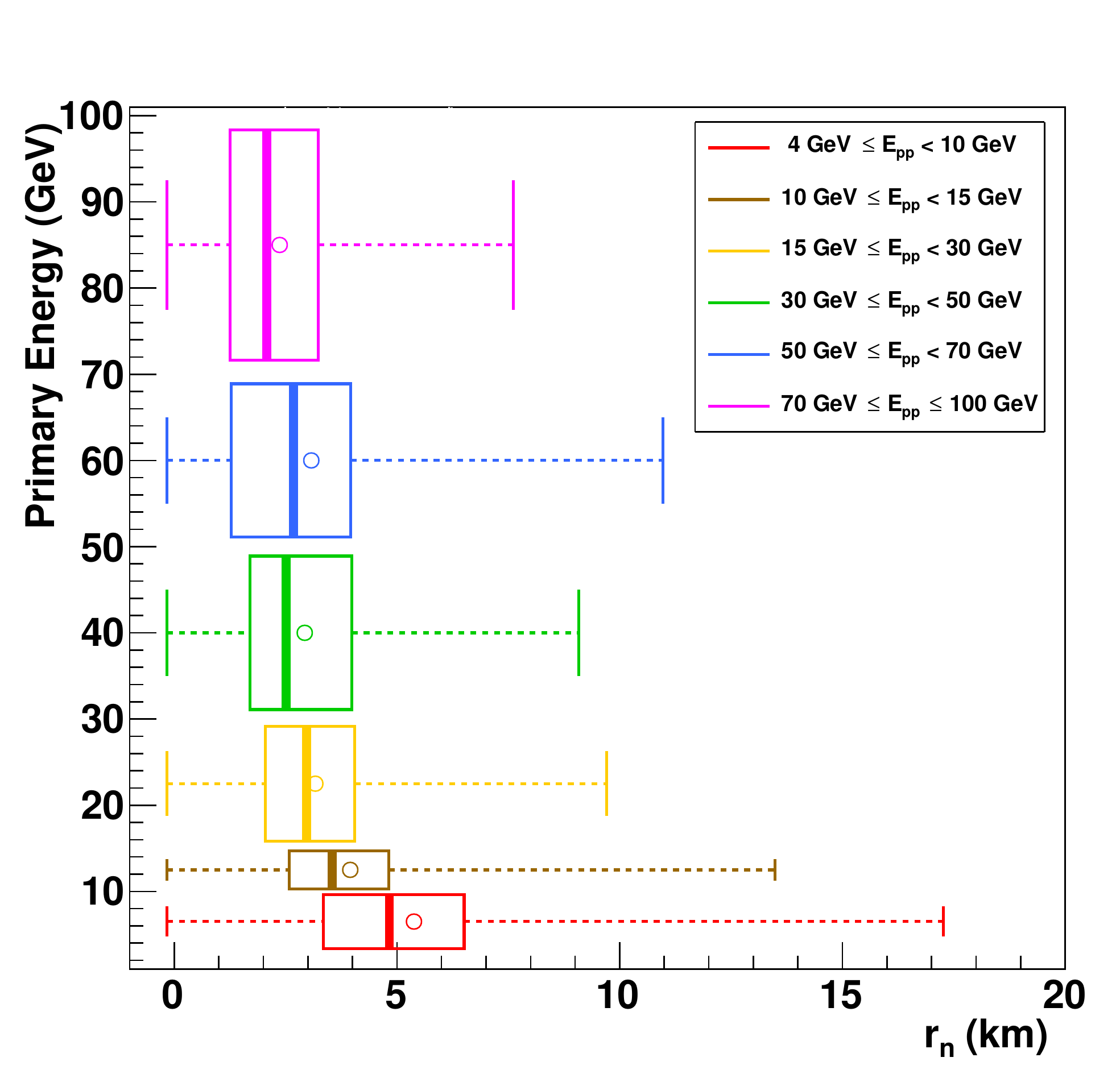} 
\caption{ (color online) Muon (top panel) and neutron (bottom panel) lateral distributions (in candle/box plot) in the polar region generated by different ranges of primary energies: 4 GeV - 10 GeV (red curve), 10 GeV - 15 GeV (brown curve), 15 GeV - 30 GeV (orange curve), 30 GeV - 50 GeV (green curve), 50 GeV - 70 GeV (blue curve), and 70 GeV - 100 GeV (magenta curve).}
\label{Rmu_NP_NoB}
\end{figure}

Another challenge of the ECRS simulation studies is to manage the simulated data and to analyze the output. This is especially true when running the ECRS simulation for different solar cycles and variable atmospheric profile to explore the long-term variations of the cosmic ray flux which could be important to for climate change studies. An example of this type of analysis is to look into the lateral distribution of the cosmic ray muons and neutrons in the polar region which are associated with the energy of the primary cosmic ray protons as shown in Fig.~\ref{Rmu_NP_NoB}.

Based on our preliminary study of ECRS simulation on XSEDE Bridges, we are highly encouraged to see that one could achieve many of the important simulation tasks to explore the applications of the cosmic rays. While continuing on optimizing the ECRS simulation software, we would like to acquire more XSEDE resource allocations to carry out focused simulations specifically associated with the global weather forecasting.  

\section{Summary and Future Work}
\label{summary}



In the paper, we briefly described the simulation software (ECRS) that have been developed at GSU for studying the cosmic ray shower characteristics in the atmosphere at global scale. We showed some of the preliminary results of running ECRS from different computing environments which include personal workstation, GSU cluster, RHIC Computing Facility, and XSEDE.  This approach closely followed the HPC model at Georgia State in that we start with local or personal resources, then scale to institutional cluster followed by national resources given the growing needs \cite{sarajlic:acore}.  The results of this simulation is of great importance with many practical applications which range from weather forecasting, muon tomography and cosmic ray related health issues.  

We see the great opportunities of accomplishing the important cosmic ray shower studies using XSEDE resources based on our initial studies for obtaining secondary cosmic ray particle distributions in the whole atmosphere, which can aid many of these important studies associated with cosmic ray applications. We also see the challenges of using XSEDE not only for carrying out ECRS simulation with high statistics but also for managing and analyzing the simulated data, which we developed workflows during our initial work via XSEDE's startup allocation. We would like to continue using the XSEDE to carry out focused simulations in near future, which we will pursue further with Research Allocation through XSEDE Resource Allocations Committee (XRAC).




\begin{acks}

We acknowledge the use of Georgia State's research computing resources that are supported by Georgia State's Research Solutions. This work used the Extreme Science and Engineering Discovery Environment (XSEDE), which is supported by National Science Foundation grant number ACI-1053575. We acknowledge the use of XSEDE resources via Campus Champion Grant - GEO150002 and Startup allocation - PHY160043.  

\end{acks}
\bibliographystyle{unsrt}%
\bibliographystyle{ACM-Reference-Format}
\bibliography{PEARC18_Sarajlic_main}

\end{document}